# The Online Resources Shared on Twitter About the #MeToo Movement:

# The Pareto Principle

**Iman Tahamtan**
University of Tennessee
Knoxville, USA
tahamtan@vols.utk.edu

**Javad Seif**
California State Polytechnic
University, Pomona, USA
jseif@ccp.edu

## ABSTRACT
In this paper we examine the most influential resources shared on Twitter about the #MeToo movement. We also examine whether a small proportion of domain names and URLs (e.g. 20%) appear in a large number of tweets (e.g. 80%) that contain #MeToo (known as the 80/20 rule or Pareto principle). R and Python were used to analyze the data. Results demonstrated that the most frequently shared domains were twitter.com (47.20%), nytimes.com (4.42%) and youtube.com (3.69%). The most frequently shared content was a recent poll which indicated "men are afraid to mentor women after the #MeToo movement". In accordance with the Pareto principle, 8% of domain names accounted for 80% of the shared content on Twitter that contained #MeToo. This study provides a base for researchers who are interested in understanding what online resources people rely on when sharing information about online social movements (e.g. #MeToo).

## ASIS&T THESAURUS
Resource sharing, Information seeking, Social media analytics

## INTRODUCTION
Social media have facilitated sharing information about different social issues. The #MeToo movement on social media, for example, encouraged women, mainly celebrities, to disclose their sexual harassment stories (Xiong, Cho, & Boatwright, 2019). Twitter is a widely used social media on which people express their stories and viewpoints by sharing URLs (Unified Resource Locators), and commenting and retweeting the tweets of other people (Java, Song, Finin, & Tseng, 2007; Stieglitz & Dang-Xuan, 2013).

Some topics, events, comments, and URLs are possibly more attractive and are accordingly shared more frequently than other topics on social media. The information sharing behavior of people on social media could probably be explained by the Pareto principle, sometimes called the 80/20 rule. According to this principle, "80 percent of the effects come from 20 percent of the causes" (Hu, Hou, & Gong, 2012, p. 724). This principle, which first appeared in the contexts of economics and engineering, has already been tested in many disciplines. For example in information science, Nicholas, Rowlands, Huntington, Jamali, and Hernández Salazar (2010, p. 430) demonstrated that this principle applies to the journal usage: "80 percent of the usage by all members coming from 20 percent of the titles".

## Research objective
In this study we investigate the most frequently shared domain names and URLs on Twitter about the #MeToo movement. We also examine whether the Pareto principle could be used to determine the most frequently shared domain names and URLs on Twitter about the #MeToo movement. Identifying the most influential resources of information can open new avenues and opportunities for further research. For example, once we know which websites have influenced a movement such as #MeToo, it is easier to see whether there has been any underlying motives or political agenda behind them. When the Pareto principle holds, we can focus on studying a small portion of these resources that are more influential than others.

## METHODS
### Data collection and cleaning
R and Python were used to retrieve and analyze the data. Twitter's Application Programming Interface (API) was used to collect tweets that included the MeToo hashtag (#MeToo). 17956 English language tweets (re-tweets were excluded from our search) were retrieved and cleaned in R in May 2019. To make sure (to a great extent) that the tweets from fake accounts and bots were excluded from our analysis, screen names that had posted multiple tweets, and the tweets with more than 5 hashtags were removed. This resulted in 10952 tweets, among which only 7175 (65.51%) contained links (URLs) to online resources.

**Data analysis**

The URLs of tweets along with their frequencies were extracted from our dataset. Twitter shows URLs in a short version which starts with http://t.co. We converted short URLs back to the original Long URLs and extracted their domain names. We then analyzed the content of the most frequently shared URLs and classified them into several categories. The main R libraries used to analyze the data were tidyverse, tidytext, lubridate, stringr, and httr.

**RESULTS**

967 URLs (out of 7175) were not found, thus were removed from our analysis. From the remaining URLs, 4832 unique URLs were identified, which have been shared 6208 times on Twitter. These URLs came from 1028 domain names. The most frequently shared domain names were twitter.com (n= 2930), nytimes.com (n=275), youtube.com (n=229), nypost.com (n=159), instagram.com (n=95), theguardian.com (n=72), npr.org (n=64), indiatimes.com (n=59), buzzfeednews.com (n=46), and washingtonpost.com (45).

The contents of the 10 most frequently shared URLs were analyzed. The most frequently shared URL (from nypost.com) was about a recent poll which indicated that "men are afraid of mentoring women after #MeToo" (n=136). Three of the 10 most frequently shared URLs (from youtube.com, cbsnews.com and breitbart.com domains) were also about this topic. We further analyzed the content of all the URLs that had been shared at least 6 times (Table 1). Similar to the results mentioned above, most URLs had covered the result of a recent poll about "men being afraid of mentoring women after #MeToo" (n=356). It was followed by "social movement against some managers in a chain restaurant that were accused of sexual harassment" (n=171), and "the responsibility of unions in both protecting from sexual harassment, and the rights of members accused of misconduct" (n=95).

| URL content | Frequency |
|---|---|
| A poll shows men are afraid to mentor women the after #MeToo movement | 356 |
| Standing against sexual harassment in a chain fast food restaurant [company name removed] in the US | 171 |
| Unions have the duty to both protect members from sexual harassment and the rights of members accused of misconduct | 95 |
| An American author and life coach [his name is removed] being accused of inappropriate sexual behaviors | 64 |
| In France, the #MeToo movement is still in its infancy | 32 |

Table 1. The most frequently shared contents on Twitter about the #MeToo movement and their frequencies

We found that domain names followed the Pareto Principle: 8% (n=84) of the domain names contributed to 80% of the shared content (which contained #MeToo) on Twitter. The Pareto Principle, however, didn't apply to URLs. 58% (n=3591) of URLs accounted for 80% of the shared content on Twitter.

**DISCUSSION**

Our result indicated that social media (e.g. twitter.com, instagram.com, and youtube.com) and news agencies websites (e.g. nytimes.com, nypost.com, theguardian.com, indiatimes.com, washingtonpost.com, cbsnews.com, usatoday.com, dailymail.co.uk and bbc.co.uk) accounted for a high percentage of the contents shared on Twitter. For example, 47.20% of the tweets contained a link to the Twitter website (twitter.com).

The most frequently shared domains and online resources about the #MeToo movement could be considered as "influencer websites" (e.g. news agencies websites). If the Pareto principle holds true for other topics as well (e.g. cyberbullying), influencer websites could be used to share credible information, and inform, update, and educate people about how to prevent and deal with the different aspects of public health and social issues. We aim to examine the Pareto principle on a larger dataset, including a wider range of topics in future studies.

**CONCLUSION**

Only a small number of domains (e.g. youtube.com, and theguardian.com) are used to share the majority of information about the #MeToo movement. This result is in line with the Pareto principle. Online resources are used to share information on Twitter about (a) "major events" (a person or company being accused of sexual harassment), (b) "major actions" that need to be taken into consideration (e.g. standing against sexual harassment), and (c) "major topics" that have received a great deal of attention (e.g. a recent poll about #MeToo).